\documentclass[letterpaper, 10 pt, conference]{ieeeconf}  

\IEEEoverridecommandlockouts                              
\usepackage[letterpaper, left=54pt, right=54pt, bottom=56pt, top=49.5pt]{geometry}

\usepackage{subcaption}
\usepackage[utf8]{inputenc}
\usepackage[fleqn]{amsmath}
\usepackage{dsfont} 
\usepackage{tabularx,booktabs}
\usepackage{units}
\usepackage{bropd} 
\usepackage{bm}    
\usepackage{currfile}
\usepackage{tikz}
\usetikzlibrary{arrows} 
\usetikzlibrary{calc} 
\usetikzlibrary{positioning}
\usepackage{pgfplots}
\pgfplotsset{compat=1.13}
\pgfplotsset{table/search path={figures//}} 
\usepgfplotslibrary{external}
\newlength\figureheight
\newlength\figurewidth
\newlength\leftlabeldist
\newlength{\bottomlabeldist}
\usepackage{myabbreviations}

\pdfoutput=1

\usepackage{hyperref}
\hypersetup{
    colorlinks=true,
    linkcolor=black,
    citecolor=black,
    filecolor=black,
    urlcolor=black,
}

\usepackage[absolute,showboxes]{textpos}

\TPMargin{5pt}

\newcommand{\copyrightstatement}{
    \begin{textblock}{0.87}(0.06,0.93)    
        \noindent
        \textcopyright 2021 IEEE.  Personal use of this material is permitted.  Permission from IEEE must be obtained for all other uses, in any current or future media, including reprinting/republishing this material for advertising or promotional purposes, creating new collective works, for resale or redistribution to servers or lists, or reuse of any copyrighted component of this work in other works.
    \end{textblock}
}

\title{\LARGE \bf
    \vspace*{-0.5pt}Optimal Trajectory Planning and Model Predictive Control of Underactuated Marine Surface Vessels using a Flatness-Based Approach
}

\author{Max Lutz and Thomas Meurer
    \thanks{Max Lutz and Thomas Meurer are with the Chair of Automatic Control, Faculty of Engineering, Kiel University, 24143 Kiel, Germany
        \mbox{\tt\small \{mlut, tm\}@tf.uni-kiel.de}}%
}

\begin{document}
\maketitle
\copyrightstatement
\thispagestyle{empty}
\pagestyle{empty}

\begin{abstract}
This paper demonstrates a refined approach to solving dynamic optimization problems for underactuated marine surface vessels.
To this end the differential flatness of a mathematical model assuming full actuation is exploited to derive an efficient representation of a finite dimensional nonlinear programming problem, which in turn is constrained to apply to the underactuated case.
It is illustrated how the properties of the flat output can be employed for the generation of an initial guess to be used in the optimization algorithm in the presence of static and dynamic obstacles.
As an example energy optimal point to point trajectory planning for a nonlinear 3 degrees of freedom dynamic model of an underactuated surface vessel is undertaken.
Input constraints, both in rate and magnitude as well as state constraints due to convex and non-convex obstacles in the area of operation are considered and simulation results for a challenging scenario are reported.
Furthermore, an extension to a trajectory tracking controller using model predictive control is made where the benefits of the flatness based direct method allow to introduce nonuniform sample times that help to realize long prediction horizons while maintaining short term accuracy and real time capability.
This is also verified in simulation where additional disturbances in the form of environmental disturbances, dynamic obstacles and parameter mismatch are introduced.
\end{abstract}
%
%
%
\section{Introduction}
With the transportation sector becoming increasingly focused on energy efficiency and low environmental impact, the research interest in automated and autonomous marine crafts has been increasing steadily.
This contribution presents a refined approach to solving the task of energy optimal point to point trajectory planning and trajectory tracking for underactuated ships taking into account static and dynamic obstacles as well as input amplitude and rate constraints.
For this an optimal control problem (OCP) is set up and transformed into a finite dimensional nonlinear programming problem (NLP) by means of a flatness based direct method exploiting the differential flatness of the considered nonlinear 3 degrees of freedom (3DOF)  dynamic vessel model.

Previous works in the context of nonlinear control of vessels include approaches using exact feedback linearization \cite{Lefeber2003}, Lyapunov-based strategies such as passivity or backstepping \cite{Fossen1999, Mazenc2002, Breivik2004, Do2005} or for the case of path following the application of model predictive control (MPC).
Some examples include the track-keeping control on inland waterways in view of rudder saturation in \cite{Wahl1998}, where the desired vessel track is represented by splines.
Approaches combining MPC with the line-of-sight concept for guidance are considered in \cite{Oh2010} and \cite{PAVLOV2009340} for underactuated marine surface vessels. 
Path following using linear MPC is proposed by \cite{Li2009} in view of roll and rudder constraints.
In contrast to path following, dynamic positioning is the task of keeping a vessel at a desired position and heading and is addressed, e.g., in \cite{Veksler2016}. 

The differential flatness property of the 3DOF vessel model considered in this work is also exploited in \cite{Helling2020}, where the flat outputs are directly parametrized using polynomial curve segments (B-splines) in order to attain a NLP for the task of vessel control in constrained areas.
The flatness based direct method used in this work however is closely related to the concept presented in \cite{Kolar2017} with the application of time optimal trajectory generation for a gantry crane, which builds on the idea of parametrizing the highest derivative of the flat output as previously suggested by, e.g., \cite{Oldenburg2002,Guay2006}. 
In those works an ansatz with basis functions for the highest derivative of the flat output is made, whereas \cite{Kolar2017} uses a piecewise constant function for this purpose.
This is also used in the context of distributed parameter systems, e.g,~by~\cite{8431201}.

The present paper adapts and extends the flatness based direct method from \cite{Kolar2017} in using continuous, piecewise linear functions as ansatz for the highest derivative of the flat output, pointing out how the significance of the flat output in original coordinates can be used to generate an initial guess for the optimization problem and combines this to realize energy optimal trajectory planning and closed loop trajectory tracking by means of optimal control and MPC, respectively.
It is organized as follows. In Sec.~\ref{sec:probFormulation} the OCP for energy optimal point to point trajectory planning is set up, the 3DOF vessel model and the approach to obstacle modeling are presented.
Section \ref{sec:numImplementation} presents the flatness based direct method used to be able to find a numerical solution to the OCP and introduces the strategy to get an initial guess.
Simulation results for an exemplary scenario follow in Sec.~\ref{sec:simRes}.
Building on this, in Sec.~\ref{sec:extMPC} the OCP is modified to allow for successive solution on a receding horizon to realize a closed loop trajectory tracking MPC.
The paper concludes with some final remarks.
%
%
%
\section{Problem Formulation}
\label{sec:probFormulation}
The task of finding an energy optimal point to point trajectory from initial time $t_0$ to final time $\te$ for a marine surface vessel can be expressed as an OCP
\begin{subequations}
    \label{eq:ocpGeneral}
    \begin{align}
        \label{eq:ocp_J}
        \min_{\uc}\; &J(\uc) = \int_{t_0}^{\te}l(t,\vec x(t),\uc(t))\dt\,, \hspace{-10ex}\\
        \label{eq:ocp_dynamic_constraint}
        \text{s.t.} \quad &\vdot x = \vec f(\vec x,\uc)\,, \quad t>t_0\,, \hspace{-10ex} &&\vec x(t_0) = \vec x_0\,, \\
        \label{eq:ocp_eqConstr}
        &\vec{g}(\vec x)=\vec{0}\,, &&\forall t\in[t_0, \te]\,,\\
        \label{eq:ocp_inEqConstr}
        &\vec{h}(\vec x, \uc, \ucDot)\leq\vec{0}\,, &&\forall t\in[t_0, \te]\,, 
    \end{align}
\end{subequations}
where the cost function \eqref{eq:ocp_J} defines the notion of optimality, \eqref{eq:ocp_dynamic_constraint} denotes the vessel dynamics  with states $\vec x(t) \in \R^n$ and inputs $\uc(t) \in \R^m$ and \eqref{eq:ocp_eqConstr}, \eqref{eq:ocp_inEqConstr} allow to introduce equality and inequality state constraints as well as constraints on the magnitude and rate of change of the input.
\subsection{Vessel Dynamics and Cost Function}
Considering near shore or harbor areas calm sea conditions can be presupposed.
This assumption allows to use a 3DOF model, neglecting roll, pitch and heave to adequately describe the vessel dynamics.
The remaining degrees of freedom given by the speeds in surge $u$, sway $v$ and the yaw rate $r$ are considered in a body fixed reference frame, commonly condensed to $\vec \nu = [u,v,r]\T$.
The vessel location and pose $\vec \eta = [x,y,\psi]\T$ are described in the inertial North-East-Down (NED) frame, where $x$ corresponds to the north, $y$ to the east position and $\psi$ to the orientation.
A commonly used nonlinear first principles model describing the equations of motion of a surface vessel is then given by
\begin{subequations}
    \label{eq:dynVesselModel}
    \begin{align}
        \vdot{\eta} &= R(\psi) \vec{\nu}\\
        M\vdot{\nu} &= -\br{C(\vec{\nu})+D(\vec{\nu})}\vec{\nu}+ \vec{\tau}\,, 
    \end{align}
\end{subequations}
see, e.g. \cite{Fossen2002}. 
Here $R(\psi)$  is the rotation matrix that transforms the body fixed velocities into the NED frame. 
The matrices $M$, $C(\vec \nu)$ and $D(\vec \nu)$ specify the inertia, coriolis and damping effects and $\vec{\tau}=[\tau_u,\tau_v,\tau_r]\T$ denotes the vector with the forces $\tau_u$ in surge, $\tau_v$ in sway direction and the yaw moment $\tau_r$ used to control the vessel.
Note that $\tau_v = 0$ for the underactuated case.
Nonlinear damping is taken into account, which yields
\begin{subequations}
\setlength\arraycolsep{1pt}
    \begin{align*}
        \nonumber
        M &= \mat{
            m_{11} & 0 & 0 \\
            0 & m_{22} & m_{23}\\
            0 & m_{32} & m_{33}}\\
            &=\mat{
            m-X_{\dot{u}} & 0 				& 0 \\
            0 			& m-Y_{\dot{v}}		& m x_g - Y_{\dot{r}} \\
            0			& m x_g - N_{\dot{v}} & I_{zz}-N_{\dot{r}}
        }\,,\\
        C(\vec{\nu}) &= \mat{
            0 & 0	& c_{13} \\
            0 & 0	& m_{11} u \\
            c_{31}	& -m_{11} u & 0
        }\,,\\
        D(\vec{\nu}) &= \mat{
            X_\txt{u}+X_{|\txt{u}|\txt{u}}|u| & 0	& 0 \\
            0 & Y_\txt{v}+Y_{|\txt{v}|\txt{v}}|v|	& Y_{\txt{r}} \\
            0	& N_{\txt{v}} & N_\txt{r}+N_{|\txt{r}|\txt{r}}|r|
        }\,,
    \end{align*}
\end{subequations}
where $m$ is the mass of the vessel, $I_{zz}$ describes its inertia about the $z$ axis of the body fixed frame, $X_{\dot{u}},Y_{\dot{v}},Y_{\dot{r}},N_{\dot{v}},N_{\dot{r}}$ are the hydrodynamic derivatives in SNAME notation and $[x_g,0,0]\T$ is the center of gravity of the ship in the body fixed frame.
Furthermore $c_{13} = -m_{22} v -\frac{m_{23}+m_{32}}{2}r$ and ${c_{31} = m_{22} v + \frac{m_{23}+m_{32}}{2}r}$\,.
With $\vec x(t) = [\vec \eta\T\!(t),\vec\nu\T\!(t)]\T \in \R^6$ and $\uc(t) = \vec \tau(t)$ the model \eqref{eq:dynVesselModel} can be written as a general nonlinear system
\begin{align}
    \label{eq:systemNonlinearGeneral}
    \vdot x &= \vec f(\vec x,\uc)\,, \quad t>t_0\,, \quad \vec x(t_0)=\vec x_0\,,
\end{align}
matching \eqref{eq:ocp_dynamic_constraint}. 
The cost function \eqref{eq:ocp_J} is designed to maintain energy efficiency with
\begin{align}
    \label{eq:costLagrangian}
    l(t,\vec x(t),\uc(t)) = \uc\!\T\!(t) Q_1 \uc(t)\,,
\end{align}
where $Q_1\in\R^{m\times m}$ is a weighting matrix, based on the assumption that the vessel energy consumption can be approximated by the time integral over the quadratic form of the actuator forces $\vec \tau = \uc$.
\subsection{Equality and Inequality Constraints}
The equality constraints \eqref{eq:ocp_eqConstr} are used to enforce agreement with initial state $\vec x_0$ and desired final state $\vec \xe$ via
\begin{align}
    \label{eq:eqConstr}
    \vec g(\vec x) &= \mat{(\vec x(0)-\vec x_0)\T &(\vec x(\te)-\vec x_\txt{e})\T}\T\,.
\end{align}
The inequality constraints \eqref{eq:ocp_inEqConstr} ensure conformity with the input constraints, as the elements of the control force and moment vector are limited both in magnitude and rate of change, i.e.,
\begin{subequations}
    \begin{align}
        \label{eq:inputConstr}
        \vec \tau_\txt{min} \leq \vec \tau \leq \vec\tau_\txt{max}\,,\\
        \label{eq:inputRateConstr}
        \vec\Delta_{\tau_\txt{min}} \leq \vdot \tau \leq \vec\Delta_{\tau_\txt{max}}\,.
    \end{align}
\end{subequations}
Obstacle avoidance can also be addressed by the inequality constraints. 
To this end the constructive solid geometry (CSG) method with smooth approximation of the intersection and union operations \cite{Ricci1973} is adopted.
It allows to efficiently include an arbitrary number of obstacles with convex or non-convex outlines.
Central to the method is the scalar valued defining function $f_{\txt{o},\,\Sigma}:\vec x \mapsto \R^+$, where $f_{\txt{o},\,\Sigma} \leq 1$ defines the area occupied by obstacles.
It is constructed by applying the approximation of the union operation
\begin{subequations}
    \begin{align}
        U_p\left(f_{\txt{o},\,1},...,f_{\txt{o},\,n}\right) &= \left(f_{\txt{o},\,1}^{-p},...,f_{\txt{o},\,n}^{-p}\right)^{-\frac{1}{p}}
        \label{eq:approxUnionCSG}
    \end{align}    
\end{subequations}
to basic geometric shapes described by the defining functions $f_{\txt{o},\,i}$, $i=1,...,n$, where the quality of the approximation increases for larger values of $p\in\N$.
Thus \eqref{eq:ocp_inEqConstr} arranges to
\begin{align}
    \label{eq:ineqConstr}
    \vec h(\vec x,\uc,\ucDot) &= \mat{\uc - \ucMax\\
                                          \ucMin-\uc\\
                                          \ucDot - \vec \Delta_{\ucMax}\\
                                          \vec \Delta_{\ucMin} - \ucDot\\
                                          1 - f_{\txt{o},\,\Sigma}(\vec x)}\,,
\end{align}
independent of the number of obstacles. 
Here $\ucMax = \vec \tau_\txt{max}$, $\ucMin = \vec \tau_\txt{min}$ and $\vec \Delta_{\ucMax} = \vec \Delta_{\tau_\txt{max}}$, $\vec \Delta_{\ucMin} = \vec \Delta_{\tau_\txt{min}}$ for the system at hand.
%
%
%
\section{Numerical Implementation}
\label{sec:numImplementation}
To be be able to use state of the art NLP solvers to efficiently solve the infinite dimensional OCP \eqref{eq:ocpGeneral} in the following a flatness based, direct method is proposed.
Exploitation of differential flatness, as introduced by \cite{Fliess1995} allows for a drastic reduction of decision variables when applying a direct method to a dynamic optimization problem, compare also \cite{Milam2000}.
As already discussed in \cite{Helling2020} the known flat parametrization of the dynamic model \eqref{eq:dynVesselModel} for the underactuated case, i.e., $\tau_v = 0$, shows singularities \cite{Agrawal2004a} such that in the following the flat parametrization of the fully actuated dynamic model is used and $\tau_v = 0$ is imposed later by setting the corresponding bounds in \eqref{eq:inputConstr} or \eqref{eq:ineqConstr}, respectively, to zero.
For the flat output $\vec z = \vec \eta$ the system states $\vec x$ and inputs $\vec \tau$ can be expressed by means of the flat parametrizations
\begin{subequations}
    \label{eq:flatParam}
    \begin{align}
        \label{eq:paramStates}
        \vec x &= \vec \theta_{\vec x}\br{\vec z,\vdot z,\dots,\vec z^{(\beta- 1)}}\,,\\
        \label{eq:paramInput}
        \vec \tau &= \vec \theta_{\vec \tau}\br{\vec z,\vdot z,\dots,\vec z^{(\beta)}}\,,
    \end{align}
\end{subequations}
with $\beta = 2$.
For the rather lengthy full expressions of the flat parametrizations for the system \eqref{eq:dynVesselModel} the interested reader is referred to \cite{Helling2020}.
Besides the benefits regarding the number of decision variables of the resulting NLP, the fact that the flat output describes the vessel position and pose in the reference coordinate system allows the generation of an initial guess for the decision variables that, despite not being completely feasible for the underactuated case, is still vastly superior over initializing with zeros or similar simple choices.
\subsection{Flatness Based Direct Method}
Generally the knowledge of a flat parametrization allows to remove the ODE constraint \eqref{eq:ocp_dynamic_constraint} from the OCP \eqref{eq:ocpGeneral} when the OCP is reformulated in terms of the flat output $\vec z$ and its derivatives $\vdot z,\dots,\vec z^{(\beta)}$, as in that case the ODE constraint is implicitly fulfilled by the flat parametrization. 
The flat parametrization also serves to reformulate the cost function \eqref{eq:ocp_J} as well as the linear \eqref{eq:ocp_eqConstr} and nonlinear constraints \eqref{eq:ineqConstr} in terms of the flat output and its derivatives up to order $\beta$.
Note that at this point the OCP is still infinite dimensional and only reformulated in the coordinates of the flat output.
The step to a finite dimensional NLP is made by using an ansatz function defined by a finite number of parameters for the highest derivative of the flat output that is required by \eqref{eq:flatParam}.
The additionally needed lower derivatives can be determined by means of successive integration without requiring smoothness of the ansatz function and are also described by the same parameters and integration constants.
These parameters and the integration constants are then used as decision variables to form a finite dimensional NLP.

To efficiently implement this concept the approach presented by \cite{Kolar2017} is build upon.
In that work $\vec z^{(\beta)}$ is parametrized as a piecewise constant step function with the step length chosen to match the discrete controller sample time. 
For the case of point to point trajectory planning for marine vessels the planning horizon $(\te-t_0)$ is magnitudes longer than the sample time of the underlying motion control system mapping $\vec \tau$ to the actual actuators, which would result in an unworkable number of parameters.
Simply increasing the step length for the parametrization of $\vec z^{(\beta)} = \vddot z$ as piecewise constant step function is clearly problematic when input rate constraints such as \eqref{eq:inputRateConstr} are taken into account:
While in principle the corresponding inequality constraints in \eqref{eq:ineqConstr} limit the change of the input values between two sample points irrespective of the ansatz function for $\vddot z$ and the discretization step size used, the system inputs will not be able to follow a step function and instead assume a piecewise linear course.
This motivates the choice for $\vddot z$ to be a piecewise linear function in this work, parametrized by the function values $\vddot z(t_k) = \vddot z_k = [\ddot z_{1,\,k}, \ddot z_{2,\,k}, \ddot z_{3,\,k}]\T$ at the points $t_k$ for $k = 0,\dots,N$ at which the linear segments are connected.

In the same manner as used in \cite{Kolar2017}, the integrator chain dynamics linking $\vec z_k$, $\vdot z_k$ and $\vddot z_k$ of the discretized problem can generally be described as discrete time systems
\begin{align}
    \begin{split}
    \label{eq:discreteTimeIntChain}
    &\vec \zeta_{i,\,k+1} = A(\TsKp)\vec \zeta_{i,\,k} + B(\TsKp)\vec u_{i,\,k}\,,\\
    &k = 0,\dots,N-1\,,\quad \vec \zeta_{i,\,0} = \mat{z_i(t_0), \dot z_i(t_0)}\T,
    \end{split}
\end{align}
for each dimension $i=1,2,3$ of the flat output, where $\vec \zeta_{i,\,k} = [z_{i,\,k},\dot z_{i,\,k}]\T$, \mbox{$\vec u_{i,\,k}=[\ddot z_{i,\,k},\ddot z_{i,\,k+1}]\T$}.
As $\vddot z$ is chosen to be piecewise linear, the dynamic matrix is given by
\begin{align*}
    A(\TsK) = \mat{1 & \TsK \\ 0 & 1}\,,
\end{align*}
and the input matrix by
\begin{align*}
    B(\TsK) = \mat{\vec b_1(\TsK) & \vec b_2(\TsK)} = \mat{\frac{\TsK^2}{3} & \frac{\TsK^2}{6} \\[0.25em] \frac{\TsK}{2} & \frac{\TsK}{2}}.
\end{align*}
Here the dependence of the matrices $ A(\TsK)$ and $B(\TsK)$ on the discretization time is stated as multiple but constant sample times $\TsOne,\dots,\TsEnd$ may be used.
The general solution of the linear discrete time system \eqref{eq:discreteTimeIntChain} is given by
\begin{align}
    \label{eq:solDTS}
    \vec \zeta_{i,\,N} &= A_N \vec \zeta_{i,\,0} + H_N\vddot z_{i,\,[0,N]}\,,  
\end{align}
where $A_N = A(\TsEnd)A(\TsEndm)\dots A(\TsOne)$ is a shorthand for successive multiplication from the left with the dynamic matrix of each discretization point and \mbox{$H_N = [\vec h_{N},\vec h_{N-1},\dots,\vec h_1,\vec h_{0}]$} with the elements 
\begin{align*}
    \vec h_{N} &= A(\TsEnd)A(\TsEndm)...A(\TsOnep)\vec b_1(\TsOne)\,,\\
    \begin{split}
        \vec h_{N-1} &= A(\TsEnd)A(\TsEndm)...A(\TsOnep)\vec b_2(\TsOne)\\
        &+ A(\TsEnd)A(\TsEndm)...A(\TsOnepp)\vec b_1(\TsOnep)\,,
    \end{split}\\
    &\vdots\\
    \vec h_1 &= A(\TsEnd)\vec b_2(\TsEndm) + \vec b_1(\TsEnd)\,,\\
    \vec h_0 &= \vec b_2(\TsEnd)\,.
\end{align*}
Finally, with $\vddot z_{i,\,[0,N]} = [\ddot z_{i,\,0},\dots,\ddot z_{i,\,N}]\T$ the composite vector of decision variables $\vec \xi \in \R^{3(N+3)}$ for the finite dimensional NLP based on the infinite dimensional OCP \eqref{eq:ocpGeneral} can be written as $\vec \xi = [z_{1,\,0},\dot z_{1,\,0},\vddot z\T_{1,\,[0,N]},\dots,z_{3,\,0},\dot z_{3,\,0},\vddot z\T_{3,\,[0,N]}]\T$.
With the algebraic equation \eqref{eq:solDTS} linking $\vec \xi$ and $\vec z_k, \vdot z_k$
Under the assumption that $\vddot z$ is piecewise linear, the OCP \eqref{eq:ocpGeneral} can now be written in the form of an NLP, i.e.,
\begin{subequations}
    \label{eq:nlpGeneral}
    \begin{alignat}{2}
        \label{eq:nlp_J}
        \min_{\vec \xi}\; &c\left(\vec \xi\right) = \sum_{k=0}^{N}l_\txt{d}\left(\vec \theta_{\vec x_k}(\vec \xi),\vec \theta_{\vec \tau_k}(\vec \xi)\right)\hspace*{-4ex}\\
        \text{s.t.} \quad &\vec{g}_\txt{d}\left(\vec \theta_{\vec x_k}(\vec \xi),\vec \theta_{\vec \tau_k}(\vec \xi)\right)=\vec{0}\,, &&k = 0, ..., N\,,\hspace*{-2ex}\\
        \label{eq:nlp_eqConstr}
        &\vec{h}_\txt{d}\left(\vec \theta_{\vec x_k}(\vec \xi),\vec \theta_{\vec \tau_k}(\vec \xi)\right)\leq\vec{0}\,, &&k = 0, ..., N\,,\hspace*{-2ex}
    \end{alignat}
\end{subequations}
with the running cost $l_\txt{d}$, equality $\vec g_\txt{d}$ and inequality constraints $\vec h_\txt{d}$ evaluated at the sample points and $\vec \theta_{\vec x_k}(\vec \xi)=\vec x_k, \vec \theta_{\vec \tau_k}(\vec \xi)=\vec \tau_k$, compare \eqref{eq:flatParam}.

A major benefit of the flatness based direct method is that there are no discretization errors as \eqref{eq:solDTS} is an exact solution at the sample points to the continuous time integrator chain dynamics linking $\vec z, \vdot z$ and $\vddot z$. 
The undisturbed system follows the predicted trajectories without error as the flat parametrizations \eqref{eq:flatParam} also are exact.
Therefore the maximum discretization time of the NLP is only limited by two aspects:
First, feasibility w.r.t. constraints is only enforced at the discretization points, and secondly, the solution is only optimal under the restriction that $\vddot z$ can be expressed by a piecewise linear, continuous function parametrized by the values at the sample points.
A formal discussion of the preservation of optimality by the flatness based direct method is beyond the scope of this contribution.
However, the underlying assumptions of a standard approach to solving the OCP \eqref{eq:ocpGeneral} using full discretization and imposing piecewise constant control inputs are essentially comparable if not~more~limiting.
\subsection{Generation of an Initial Guess}
As the OCP \eqref{eq:ocpGeneral} and thus also the NLP derived from it by means of the flatness based direct method outlined before are highly nonlinear and non-convex the importance of a good initial guess for a fast and meaningful solution can not be overrated. 
To this end it is exploited that the flat output $\vec z$ for the system \eqref{eq:dynVesselModel} is the vessel position and pose $\vec \eta = [x,y,\psi]\T$ in the inertial frame.
There are highly efficient algorithms to solve the point to point path finding problem on a grid in the presence of obstacles such as the \aStar-algorithm \cite{Hart1968}.
With $\beta = 2$ in \eqref{eq:flatParam} however, any initial guess for the flat output $\vec z$ needs to be twice continuously differentiable.
This can be accomplished by convoluting the non-smooth path provided by the \aStar-algorithm with a mollifying function, also called mollifier. 
A 1D function $\phi(t)$ on $\R^1$ is a mollifier if it exhibits the properties: (i) it is compactly supported, (ii) $\int_\R \phi(t)\dt = 1$ and (iii) \mbox{$\lim _{\epsilon \to 0}\phi _{\epsilon }(t)=\lim _{\epsilon \to 0}\epsilon ^{-1}\phi (t/\epsilon )=\delta (t)$}, where $\delta(t)$ is the Dirac delta function.
The details of generating an initial guess in a multi-stage approach of path finding, simplifying the path, adding timing information, smoothing the resulting trajectory and mapping it to the decision variables will be outlined in the following.
It has to be pointed out that the resulting initial guess is not necessarily fully feasible w.r.t. the input constraints \eqref{eq:inputConstr} and \eqref{eq:inputRateConstr}, including $\tau_v = 0$.
However, the convergence times of the solver for the underactuated case are still greatly reduced when compared to a more trivial initial guess such as, e.g., initialization with zeros.

\begin{figure}
    \vspace*{5pt}
    \centering{
        \includegraphics{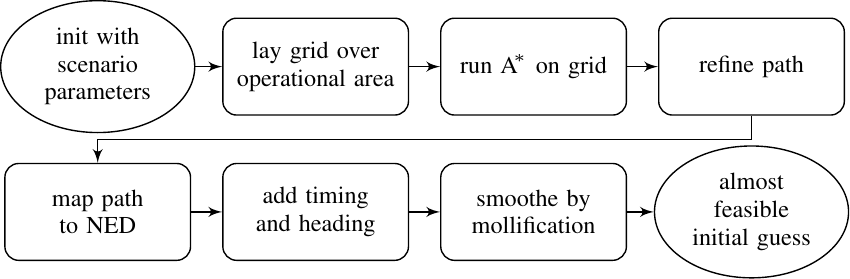}
    }
    \vspace*{-12.5pt}
    \caption{Flow chart of algorithm used to generate initial guess.}
    \label{fig:initGuessFlow}
\end{figure}

Figure \ref{fig:initGuessFlow} illustrates the steps undertaken to generate the initial guess.
The \aStar-algorithm operates on a grid and requires the specification of a heuristic function that expresses the expected ideal path length from the current node to the end node. 
A rectangular grid is laid over the area of operation of the vessel and start and end node are chosen to be the nodes that are closest to the actual desired initial and end position and not within an obstacle.
Allowed node transitions are such that each node is connected to its eight neighboring nodes in the rectangular grid and obstacle information is provided by means of the defining function $f_{\txt{o},\,\Sigma}$, compare \eqref{eq:ineqConstr}.
In a subsequent step the resulting path is refined by first removing all path nodes that are not directly adjacent to an obstacle.
It is then transformed in the NED frame by remapping the path grid points to points in the NED frame.
Lastly the start and end node position are replaced with the actual initial and desired final position. 
Timing information is added by calculating a theoretical average along-track speed based on the path length and the optimization time $\te$.
From the waypoints and the constant along track speed piecewise constant curves $\dtilde x = \dtilde z_1$ and $\dtilde y = \dtilde z_2$ can be determined.
For the heading it is desired that the vessel is always oriented parallel to the path segment, leading to a sequence of impulses $\dtilde \psi = \dtilde z_3$, where the time integral of each impulse equals the change in path angle at that waypoint.

As the decision variables of the NLP are the second derivatives of the flat output the differentiation property of the convolution $\od{}{t}\br{f*g} = \od{f}{t}*g=f*\od{g}{t}$ for functions $f(t), g(t)$ comes handy.
Accordingly, the initial guess for the decision variables $\vec \xi$ is generated by sampling the convolution $\dtilde z_i*\dot \phi_i$, $i=1,2,3$ at the discretization points of the NLP.
Here $ \dot \phi_i(t)$
is the time derivative of the mollifier
\begin{align*}
    \phi_i(t) &= \begin{cases}
        \frac{15}{16\epsilon_i}\br{1-\frac{t}{\epsilon_i}^2}^2\;, &\text{for} -\epsilon_i \leq t \leq \epsilon_i\,,\\
        0\,, & \text{else},                         
    \end{cases}
\end{align*}
where $\epsilon_i$ is a tuning parameter controlling the amount of smoothing with $\lim_{\epsilon \to 0}\br{\dtilde z_i*\dot \phi_i} =\dtilde z_i$, $i=1,2,3$.
Initial and final speeds $\vec \nu_0$, $\vec \nu_\txt{e}$ can be taken into account by transforming them into the NED frame with $\vdot z_0^* = R(\psi_0)\vec \nu_0$, $\vdot z_\txt{e}^* = R(\psi_\txt{e})\vec \nu_\txt{e}$ and then extending the signals $\dtilde z_i$ defined on $t \in [t_0,\te]$ by mirroring them at the points $(t_0,z^*_{0,\,i})$ and $(\te,z^*_{\txt{e},\,i})$, $i = 1,2,3$\,.
With this approach an initial guess is obtained that fulfills the ODE constraint \eqref{eq:ocp_dynamic_constraint} via the flat parametrization.
Ideally there is no violation of the equality constraints \eqref{eq:eqConstr} and the inequality constraints \eqref{eq:ineqConstr} except those concerning the input.
A possible slight violation can be caused by the \aStar-algorithm operating on a discrete grid, the smoothing process and the sampling of the smoothed signal.
Generally, for larger values of $\epsilon_i$ and thus increased smoothing there is a larger distortion of the path compared to the output of \aStar and thus a higher possibility of violating state constraints, but also the extrema of $\vddot z$ are smaller and thus better agreement with the input constraints is achieved. 
%
%
%
\section{Simulation Results}
\label{sec:simRes}
To allow for a future verification of the theoretical results of this paper in a laboratory scale setup, the parametrization of a model vessel from \cite{Do2006} is used as displayed in Table~\ref{tab:parameters}. 
\begin{table}[!t]
    \vspace{3pt}
    \centering
    \caption{Basic vessel parameters.}
    \label{tab:parameters}
    \begin{tabular}{lcrlclcrl} 
        $m_{11}$ &\hspace{0.1cm} &25.8 & $\unit{kg}$& \hspace*{4.5ex}&$Xu$    &\hspace{0.1cm} & 12.0  &$\unitfrac{kg}{s}$\\ 
        $m_{22}$ & &33.8 & $\unit{kg}$ 			 	& &$Yv$    & & 17.0  &$\unitfrac{kg}{s}$\\
        $m_{23}$ & &6.2  & $\unit{kg}\,\unit{m}$ 	& &$Yr$    & & 0.2   &$\unit{kg}\,\unitfrac{m}{s}$ \\
        $m_{32}$ & &6.2  & $\unit{kg}\,\unit{m}$  & &$Nv$    & & 0.5   &$\unit{kg}\,\unitfrac{m}{s}$ \\
        $m_{33}$ & &2.76 & $\unit{kg}\,\unit{m^2}$ & &$Nr$    & & 0.5   &$\unit{kg}\,\unitfrac{m^2}{s}$\\
        length& &1.2 &\unit{m}& &$X_{|u|u}$ &\hspace{0.1cm} & 2.5   &$\unitfrac{kg}{m}$ \\	
        width& &0.35 &\unit{m} & &$Y_{|v|v}$ & & 4.5   &$\unitfrac{kg}{m}$ \\ 
        mass& &17 &\unit{kg} & & $N_{|r|r}$ & & 0.1   &$\unit{kg}\,\unit{m^2}$
    \end{tabular}
\end{table}
The inputs are constrained to $\unit[-5]{N}\leq \tau_u \leq \unit[5]{N}$, $\unit[-0.2]{Nm}\leq \tau_r \leq \unit[0.2]{Nm}$ and $0\leq \tau_v \leq 0$ as the underactuated case is considered. 
Further the input rate limits are chosen to be $\unitfrac[-0.5]{N}{s}\leq \dot\tau_u \leq \unitfrac[0.5]{N}{s}$ and $\unitfrac[-0.1]{Nm}{s}\leq \dot\tau_r \leq \unitfrac[0.1]{Nm}{s}$.

The scenario is constructed such that the vessel has to navigate a narrow channel between two obstacles.
The obstacle defining function is constructed by applying the approximate union \eqref{eq:approxUnionCSG} with $p=5$ to four basic shapes of the form
\begin{multline}
    \label{eq:csg_rectangle}
    f_{\txt{o}}(\vec x) = \Bigg[\bigg(\frac{2\br{\cos\alpha(x-x_{\txt{o}}) + \sin\alpha(y-y_{\txt{o}})}}{\delta_{x}} \bigg)^{2a}\\
    +\bigg(\frac{2\br{-\sin\alpha(x-x_{\txt{o}})+\cos\alpha(y-y_{\txt{o}})}}{\delta_{y}} \bigg)^{2a}\Bigg]^{\frac{1}{a}}\,.
\end{multline}
The parameters $x_{\txt{o}},\, y_{\txt{o}},\, \delta_{x},\, \delta_{y},\, \alpha$ describe the center position, length,
width and orientation of the shape in the NED reference frame and the parameter $a\in \N$ determines the rounding of the corners.
For $a = 1$ the shape is circular and for $a \rightarrow \infty$ it becomes rectangular.
The parameters of the basic shapes used are summarized in Table~\ref{tab:basicShapes}.
    \begin{table}[!t]
        \centering
        \caption{Basic obstacle shapes.}
        \label{tab:basicShapes}
        \begin{tabular}{ccccccc} 
            $i$     & $x_{\txt{o}}$ (m) & $y_{\txt{o}}$ (m) & $\delta_{x}$ (m) & $\delta_{y}$ (m) & $\alpha$ ($^\circ$)& $a$\\
            \midrule
            1&6.5&14&1&2.5&0&2\\
            2&1&15&1&2.5&0&3\\
            3&6&8&5&2&-15&1\\
            4&-1&18&8&1&-10&1
        \end{tabular}
    \end{table}
The initial state is $\vec x_0 = [0,0,\nicefrac{\pi}{2},0,0,0]\T$ with $\vec \tau_0 = [0,0,0]\T$ and the desired terminal state is $\vec x_\txt{e} = [1,30,\nicefrac{\pi}{2},0,0,0]\T$ for $t_0 = 0$ and $t_\txt{e} = \unit[120]{s}$.
The weight matrix $Q_1$ in the running cost \eqref{eq:costLagrangian} is set to $Q_1 = \text{diag}\br{(\nicefrac{1}{\tau_{u,\,\txt{max}}})^2,\, 0,\, (\nicefrac{1}{\tau_{r,\,\txt{max}}})^2}$ and the cost function is integrated via trapezoidal quadrature.
For the initial guess a 20$\times$40 grid is put over the map area $(\unit[-1]{m}\leq x \leq \unit[9]{m}$, $\unit[-1]{m}\leq y \leq \unit[31]{m})$ to which the \mbox{\aStar-algorithm} is applied.
The tuning parameters of the mollifiers are set to $\epsilon_1 = \epsilon_2 = 0.5$ for the smoothing of the initial guess of the position and $\epsilon_3 = 1.6$\,.

\begin{figure}[!t]
    \vspace{2pt}
    \captionsetup[subfigure]{aboveskip=-1pt,belowskip=0pt}
    \begin{subfigure}{\linewidth}
    \centering{
        \includegraphics{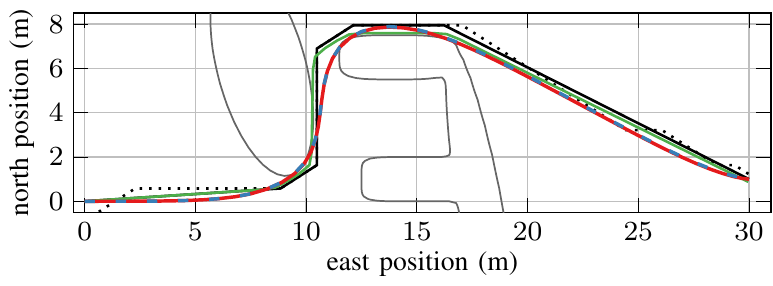}}
    \captionsetup{width=\linewidth}
    \subcaption{Path in the NED-Frame with obstacles (gray), \aStar path (dotted), refined path (black), initial guess (green), OCP output (red) and simulation output (blue, dashed).} 
    \label{fig:ocpXY}
    \end{subfigure}
    \begin{subfigure}{\linewidth}
    \centering{
        \includegraphics{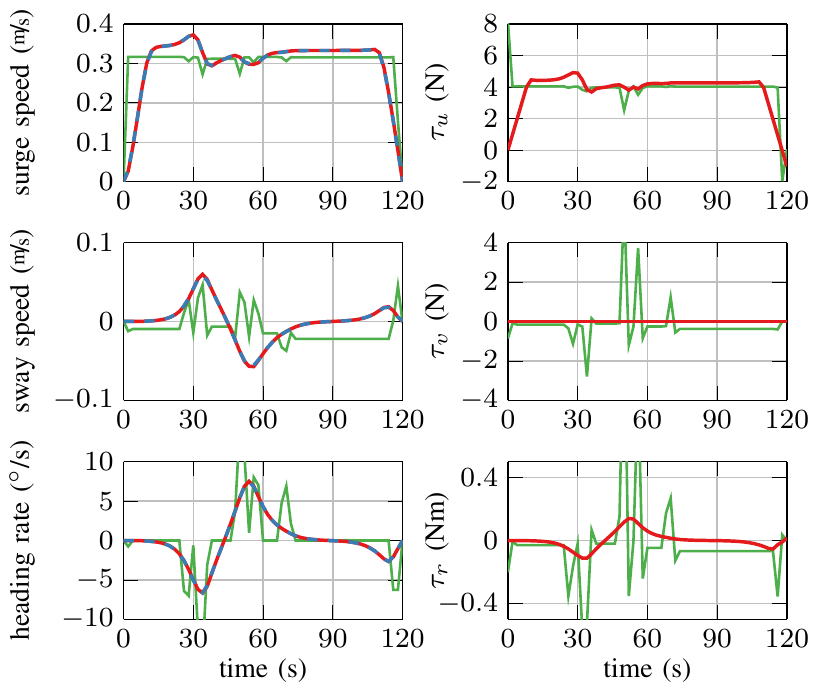}}
    \subcaption{Initial guess (green), OCP output (red) and simulation results (blue) for the body fixed speeds (left) and corresponding force and moment control inputs (right).}
    \label{fig:ocpNu}
    \end{subfigure}
    \caption{Initial guess, optimization output and simulation results for the energy optimal OCP with discretization time $\Ts = \unit[2]{s}$.}
\end{figure}

The NLP is programmed in \matlab and solved using the \matlab-interface of SNOPT \cite{GilMS05} with a discretization time of $\Ts = \unit[2]{s}$ equaling 61 discretization points or 189 decision variables.
On a normal laptop computer with Intel Core i5-6200U CPU clocked at \unit[2.30]{GHz} the optimization takes on average about \unit[1.5]{s}, whereas for an initial guess of zeros for the exact same problem the solver fails to find the opening between the obstacles and does not converge within a time limit of \unit[1000]{s}.
The vessel equations are simulated in original coordinates \eqref{eq:dynVesselModel} using \matlab's \texttt{ode45()} variable step ODE solver with the inputs calculated by the NLP solver.
Figure~\ref{fig:ocpXY} shows the resulting trajectory in the NED frame and Fig. \ref{fig:ocpNu} displays the corresponding speeds in the body fixed frame as well as the trajectories of the components of the force and moment vector $\vec \tau$.
The discerning reader may notice that in some intervals the trajectory is planned right on the border of the obstacles. 
This is a direct consequence of the energy optimality requirement, in much the same way as a race car clips the inside of a corner to preserve momentum.
To illustrate the effectiveness of the proposed approach in attaining an energy optimal trajectory, it is compared to the solution using a running cost penalizing path length and speed variations,
\begin{align}
    \label{eq:shortestPathSteadySpeedLagrangian}
    l(t,\vec x(t),\uc(t)) = \sqrt{x(t)^2+y(t)^2} + c_1(t)\dot\tau_u^2(t)\,.
\end{align}
Here the penalty parameter is chosen such that the initial acceleration and final deceleration are not penalized, i.e., ${c_1(t<\unit[10]{s}\vee t>\unit[110]{s})} = 0$ and ${c_1(\unit[10]{s}\leq t\leq \unit[110]{s})} = 10$.
The results of a cross-evaluation of the energy and distance measure, given by the integral over \eqref{eq:costLagrangian} and \eqref{eq:shortestPathSteadySpeedLagrangian} with $c_1(t) = 0,\,\forall t$, are given in Table~\ref{tab:performance}.
While the path lengths differ only slightly, the difference in actuator energy cost is significant.
\begin{table}[!t]
    \centering
    \vspace{4pt}
    \caption{OCP cost function performance comparison.}
    \label{tab:performance}
    \begin{tabular}{llrr} 
        && \hspace{-7ex}energy measure & distance\\ 
        \midrule
        \multirow{2}{*}{OCP}&energy minimal running cost \eqref{eq:costLagrangian} &$\bm{85.3}$&\unit[36.3]{m}\\
        &shortest distance running cost \eqref{eq:shortestPathSteadySpeedLagrangian}&100.3&$\unit[\bm{35.8}]{m}$\\
    \end{tabular}
\end{table}
%
%
%
%

\section{Extension to MPC}
\label{sec:extMPC}
The availability of a good initial guess and the lack of discretization errors are especially useful in the design of a closed loop trajectory tracking controller based on MPC dealing with disturbances such as non-cooperative dynamic obstacles.
Real time capability is crucial in this context, which limits the number of discretization points.
Typically this results in a trade-off between high accuracy using short sample times and long prediction horizons that require longer sample times.
As the flatness based direct method does not introduce discretization errors even though the system \eqref{eq:dynVesselModel} is nonlinear it is possible to diminish this concession by using multiple sample times.
Here three sample times $\TsOne < \TsOnep < \TsOnepp$ for a total number of $N = 1 + N_1 + N_2 + N_3$ discretization points yield a horizon length of $\tHor = N_1\TsOne + N_2\TsOnep + N_3\TsOnepp$. 
This allows to choose $\TsOne$ small such that the control is updated at a high rate, $\TsOnep$ such that near future prediction is sufficiently accurate and $\TsOnepp$ large to allow long prediction times maintaining a reasonable number of discretization points.
Often in MPC the result of a previous iteration is used as the basis for the initial guess of the next iteration. 
The issue with this approach in the case of non-convex problems -- as the one at hand -- can be illustrated with the example of an adversarial dynamic obstacle.
To this end the dynamic obstacle is stationary on the reference trajectory such that initially when it appears on the MPC prediction horizon the controller plans to avoid it by crossing ahead.
However, then the obstacle starts to move, making it advantageous to cross behind.
Even with the now changed situation the use of the previous result as an initial guess for the next MPC iteration typically leads to the controlled vessel attempting to cross ahead, as the optimization keeps converging to the same local minimum. 
The generation of an initial guess independently from the previous result of the MPC does not suffer from this memory effect as will be illustrated in the following.

The successful extension of the OCP to form a MPC by repeatedly solving the OCP on a receding time horizon requires some adaptations to \eqref{eq:ocpGeneral}.
First, the inequality constraint in \eqref{eq:ocp_inEqConstr} corresponding to the obstacle constraints is rewritten as a slack constraint $\vec h(\vec x, \uc, \ucDot) - [\vec 0\T s]\T \leq \vec 0\,$, where $s \in \R_{0^+}$ is the slack variable.
This has the benefit that in the case of constraint violation due to external disturbances the problem does not become infeasible.
The cost function \eqref{eq:ocp_J} is extended to
\begin{align}
    \label{eq:closestPointEnergy}
    \begin{split}
        J(\uc) &= \int_{t_0}^{\tHor}\uc\!\T\!(t) Q_1 \uc(t)\dt + q_2 s^2 + q_3 s\\
                &+ \vec \Delta_{\vec{x}}\!\T(\tHor)Q_4 \vec \Delta_{\vec{x}}(\tHor)\,,
    \end{split}
\end{align}
with the weights $q_2, q_3$ and weighting matrices $Q_1, Q_4$ and $\vec \Delta_{\vec{x}}(t) = \vec x(t) - \vec x^*(t)$, where $\vec x^*(t)$ denotes the reference trajectory.
This cost function realizes a last-waypoint-match strategy (LWM), where the vessel can plan a locally energy optimal path, however a deviation of the predicted state at the end of the MPC horizon $\vec x(\tHor)$ from the corresponding desired state as specified by the reference trajectory $\vec x^*(\tHor)$ is penalized.
The weights $q_2, q_3 \in \R_{0^+}$ allow to control the behavior w.r.t. the slack constraints, roughly speaking the former mainly controls how significantly and the latter for how long constraints may be violated.

\begin{figure}[!t]
    \vspace{3pt}
    \captionsetup[subfigure]{aboveskip=-1pt,belowskip=0pt}
    \begin{subfigure}{\linewidth}
        \centering{
            \includegraphics{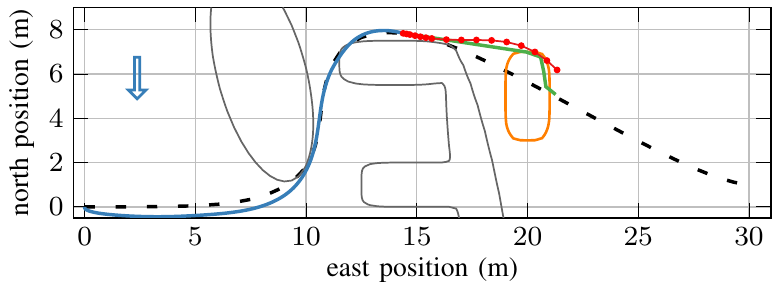}}
        \captionsetup{width=\linewidth}
        \subcaption{NED-frame at $t_0 = \unit[65]{s}$, dynamic obstacle is stationary.} 
        \label{fig:mpcXYAhead}
    \end{subfigure}
    \begin{subfigure}{\linewidth}
        \centering{
            \includegraphics{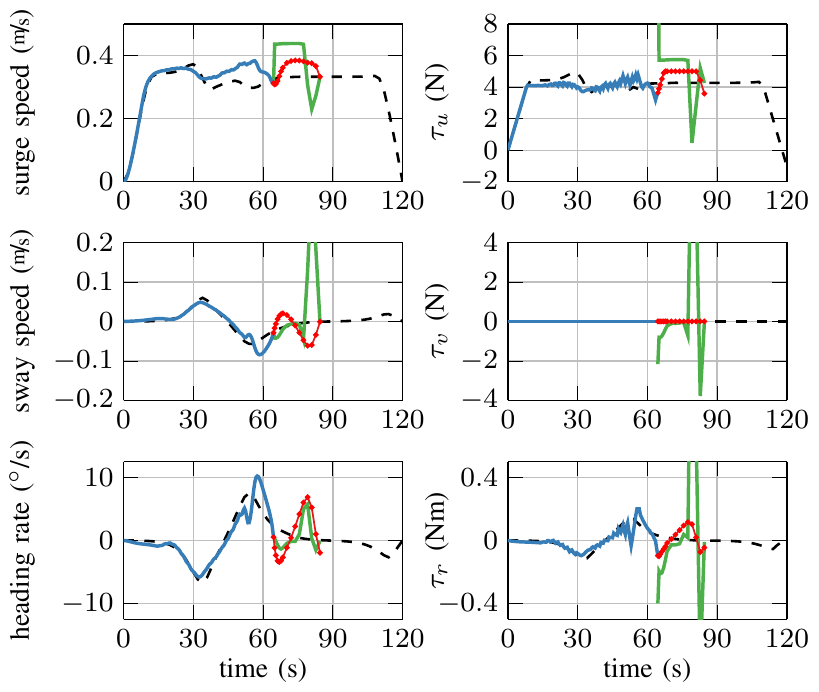}}
        \captionsetup{width=\linewidth}
        \subcaption{Body fixed speeds (left), control inputs (right) at $t = \unit[65]{s}$.} 
        \label{fig:mpcNuAhead}
    \end{subfigure}
    \begin{subfigure}{\linewidth}
        \centering{
            \includegraphics{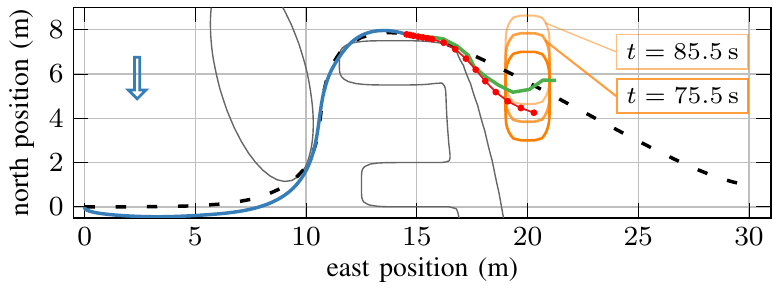}}
        \captionsetup{width=\linewidth}
        \subcaption{NED-frame at $t_0 = \unit[65.5]{s}$, the MPC reacts to changed situation}
        \label{fig:mpcXYBehind}
    \end{subfigure}
    \caption{Simulation Results for the MPC with reference trajectory (dashed), simulation output (blue), MPC prediction (red), static obstacles (gray), dynamic obstacle (orange) and initial guess (green). The arrow indicates the ocean current.}
    \vspace*{-0.25cm}
\end{figure}

The MPC is also implemented using the flatness based direct method reported above, with the simulation result from Sec.~\ref{sec:simRes} serving as a reference trajectory.
The general scenario and parameters are maintained, however to introduce disturbance the vessel is simulated using \matlab's \texttt{ode45()} with a $\unit[-10]{\%}$ parameter mismatch w.r.t. all parameters of the MPC prediction model.
Motivated by \cite{BITAR2018389}, a north-to-south ocean current with a magnitude of \unitfrac[0.04]{m}{s}, which is about \unit[10]{\%} of the vessels maximum speed, serves to demonstrate the effectiveness of the obstacle constraints with slack.
As disturbance estimation is beyond the scope of this contribution no disturbance estimator is used, thus this constitutes a severe disturbance.
Finally an adversarial dynamic obstacle with a time dependent position $(x_\txt{o,\,dyn} \; y_\txt{o,\,dyn}) = {(5\; 20)\,, t<65}$, $(x_\txt{o,\,dyn} \; y_\txt{o,\,dyn}) = (5\;20) + (0.08\;0)(t-65)\,,t>65$, all in meters or seconds, respectively, is introduced.
Further parameters that are not present in the OCP setting are chosen as: $q_2 = 1\pTen{3}$, $q_3 = 1\pTen{2}$, $\tHor = \unit[20]{s}$, $\TsOne = \unit[0.5]{s}$, $\TsOnep = \unit[0.75]{s}$, $\TsOnepp = \unit[1.78]{s}$ with $N_1 = 2$, $N_2 = 4$ and $N_3 = 9$.
The simulation results show that the controller can limit the effect of the disturbance and that the penetration of the obstacles remains minimal.
Feasibility is not lost at any point, with average optimization times of \unit[250]{ms} on the reported hardware.
Figures \ref{fig:mpcXYAhead} and \ref{fig:mpcXYBehind} show how due to the \aStar based initial guess the MPC reacts to the actions of the adversarial dynamic obstacle in the previously mentioned scenario within one loop iteration, Fig.~\ref{fig:mpcNuAhead} shows body fixed velocities and control inputs for the encounter.
Applying the same performance measures as in the OCP case, when compared to an all-waypoint-match type running cost, i.e.,
\begin{align}
    \label{eq:allWaypointMatch}
    l(t,\vec x(t),\uc(t)) = \vec \Delta_{\vec{x}}\!\T(t)Q \vec \Delta_{\vec{x}}(t)\,
\end{align}
with $Q = 100\cdot\text{diag}\left(1,1,1\right)$, the energy minimal LWM strategy again shows lower energy costs, see Table~\ref{tab:performanceMPC}.
Note that the values for the performance measures of OCP and MPC can not be compared as for the latter the vessel is simulated with the parameter mismatch and the influence of the current.
\begin{table}[!t]
    \vspace{4pt}
    \centering
    \caption{MPC cost function performance comparison.}
    \label{tab:performanceMPC}
    \begin{tabular}{llrr} 
        && \hspace{-7ex}energy measure & distance\\ 
        \midrule
        \multirow{2}{*}{MPC}&energy minimal LWM \eqref{eq:closestPointEnergy} &$\bm{79.5}$&\unit[37.2]{m}\\
        &all-waypoint-match \eqref{eq:allWaypointMatch}&85.6&$\unit[\bm{36.9}]{m}$\\
    \end{tabular}
\end{table}
\addtolength{\textheight}{-0.7cm}   
%
%
%
\section{Conclusion}
The energy optimal point to point trajectory planning for underactuated autonomous marine surface vessels is addressed using nonlinear optimal control with an extension to trajectory tracking using model predictive control.
For this the differential flatness of a 3DOF model of the vessel dynamics with input constraints, both in magnitude and rate, as imposed, e.g., by the propulsion and rudder subsystem is exploited by means of a flatness based direct method in which the highest derivative of the flat output is specified to be a piecewise linear, continuous function.
Furthermore the meaning of the flat output in original coordinates together with the properties of the flatness based direct method are used to develop a novel approach to generate an initial guess for the optimization algorithm.
Here the smoothing properties of a mollifier are used to obtain sufficiently smooth functions based on an \aStar solution to the point to point constrained path finding problem on a discrete grid.
Simulation results support the theoretical development in a scenario with multiple static obstacles for the open loop optimal control and an additional disturbance due to a non-cooperative dynamic obstacle, a parameter mismatch and the effect of ocean current for the closed loop model predictive control.
As a next step the implementation of the presented schemes on a laboratory scale test vessel will be performed.

\bibliographystyle{IEEEtran}
\bibliography{mpcFlatnessBottomUP_lit}

\end{document}